\documentclass{elsart}

\usepackage{natbib}
\usepackage{graphicx}

\usepackage{amssymb}
\usepackage{amsmath}
\usepackage{bm}

\def\url#1{{\ttfamily\def\/{/\discretionary{}{}{}}#1}}
\def\bibcode#1{}

\begin{document}

\begin{frontmatter}
\title{Reducing the shear}
\author[address1,address2]{Martin White\thanksref{mwemail}},
\address[address1]{Department of Astronomy, University of California,
Berkeley, CA, 94720}
\address[address2]{Department of Physics, University of California,
Berkeley, CA, 94720}
\thanks[mwemail]{E-mail: mwhite@berkeley.edu}

\begin{abstract}
As gravitational lensing measurements become increasingly precise, it
becomes necessary to include ever higher order effects in the theoretical
calculations.  Here we show how the difference between the shear and
the reduced shear manifest themselves in a number of commonly used measures
of shear power.
If we are to reap the science rewards of future, high precision measurements
of cosmic shear we will need to include this effect in our theoretical
predictions.
\end{abstract}

\begin{keyword}
Cosmology \sep Lensing \sep Large-Scale structures
\PACS 98.65.Dx \sep 98.80.Es \sep 98.70.Vc
\end{keyword}
\end{frontmatter}

\section{Introduction}

Weak gravitational lensing by large-scale structure
\citep{LensReview1,LensReview2}
is becoming a central means of constraining our cosmological model
\citep[see e.g.][for the current status]{HoeYeeGla,WaeMelHoe}.
Continuing effort has yielded increasingly stringent control of
systematic errors and ever larger surveys are decreasing the
statistical errors rapidly.
As the precision with which the measurements are made increases,
one demands ever higher fidelity in the data-analysis, modeling and
theoretical interpretation.

One such area is the computation of the theoretical predictions for
various statistics involving the measurable `reduced shear'
\citep{LensReview1,LensReview2}
\begin{equation}
  g \equiv \frac{\gamma}{1-\kappa}\ {\rm for}\ |g|<1
\label{eqn:red}
\end{equation}
where $\gamma$ is the `plain' shear and $\kappa$ is the convergence
which is related to the projected mass along the line of sight.
Both $\kappa$ and $\gamma$ are defined through the Jacobian of the
mapping between the source and image planes.
If the lensing is weak then $|\kappa|\ll 1$, and to lowest order
$g=\gamma$ and the power spectra of the shear and convergence are
equal: $C_\ell^\gamma=C_\ell^\kappa$.
Thus measurements are usually compared to predictions of $C_\ell^\kappa$
derived from models of the non-linear clustering of matter.
This relation has corrections however, which need to be taken into
account if precise comparisons with data are to be made in the future.
While there has been previous analytic work on this subject
\citep{Map,DodZha}
and some numerical work on magnification statistics
\citep{TakHam,BarTay,MHBY},
there has been no comprehensive investigation of this effect in simulations.
In this short paper we present some results, derived from simulations
similar to those described in \citep{LensGrid}, on the size of the
corrections for a variety of commonly used statistics on the angular scales
where most current lensing work has focused.

\section{Simulation}

The results are derived from 7 N-body simulations of a $\Lambda$CDM
cosmology run with the TreePM code \citep{TreePM}.  Each simulation
employed $384^3$ equal mass dark matter particles in a periodic
cubical box of side $200\,h^{-1}$Mpc run to $z=0$ with phase space
data dumped every $50\,h^{-1}$Mpc starting at $z=4$.  The cosmology
was the same for each simulation ($\Omega_{\rm mat}=0.28$,
$\Omega_Bh^2=0.024$, $h=0.7$, $n=1$ and $\sigma_8=0.9$) but different
random number seeds were used to generate the initial conditions.
The purpose of these runs was to allow a study of the statistical
distributions of lensing observables.
Each box was used to generate 16 approximately independent lensing maps,
each $3^\circ\times 3^\circ$, using the multi-plane ray-tracing code
described in \citep{ValWhi}.  Each map was produced with $2048^2$ pixels
and then downsampled to $1024^2$ pixels\footnote{Tests with maps downsampled
to $512^2$ pixels indicate that our results are converged at the several
percent level in the ratios that we quote, with better agreement at larger
scales as expected.}.
Thus a total of 112 maps or $1008$ square degrees were simulated, all with
the same cosmology.
We report on the results for sources at $z\simeq 1$ (specifically all at
a comoving distance of $2400\,h^{-1}$Mpc) here.
Other data can be obtained from http://mwhite.berkeley.edu/Lensing.

\section{Results}

\subsection{Two point statistics}

The lowest order information on large-scale structure comes from studying
the 2-point statistics of the shear field.  A number of different 2-point
statistics have been used in the literature, each with their own strengths
and weaknesses.  We survey the most commonly used statistics here.

\begin{figure}
\begin{center}
\resizebox{5.5in}{!}{\includegraphics{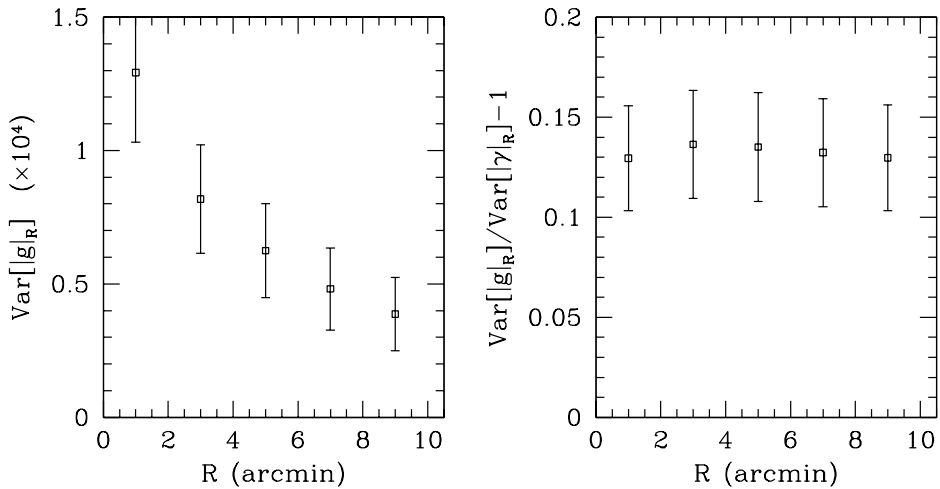}}
\end{center}
\caption{(Left) The smoothed shear variance, Var$[|g|_R]$, as a function
of smoothing scale $R$.
(Right) The relative difference of the smoothed shear variance computed
for plain shear and reduced shear, as a function of smoothing scale $R$.
The points show the mean difference, averaged over 112 maps, each
$3^\circ\times 3^\circ$, while the error bars indicate the standard deviation.}
\label{fig:svar}
\end{figure}

We first show the results on the variance of the shear, smoothed on
a range of angular scales.  Though it has numerous drawbacks in
interpretation, this was one of the first statistics computed from
shear maps due to its ease of computation.  The left panel of
Figure \ref{fig:svar} shows the (reduced) shear variance as a function
of smoothing scale.  For each point we compute a map of the amplitude of
the reduced shear, smooth the map with a 2D boxcar of side length $R$ and
then compute the variance of the resulting map.
The points are highly correlated.
Since we have full information about the shear and convergence from the
simulation we are able to compare this with the plain shear results which
are usually computed.
The right panel of Figure \ref{fig:svar} shows the difference of this
statistic computed for the reduced shear ($g$) to that for plain shear
($\gamma$).
We see that the reduced shear variance is larger than the usually predicted
shear variance by a non-trivial amount.  The bias very gradually drops as we
go to larger scales (not shown here) but we are unable to follow it to
convergence due to the finite size of the fields simulated.

Naively one might imagine that the corrections would be smaller than shown
in Figure \ref{fig:svar} because $\kappa$ is small `on large scales'.
However $\kappa$ is only small when averaged over a sizeable region of sky,
and this smoothing does not commute with the division in Eq.~(\ref{eqn:red}).
This fact also means that perturbative calculations must be used with care,
because the convergence or shear amplitude can be quite large on small
scales leading to a break down of the approximation.

Though the smoothed shear variance has the worst behavior of the two point
functions we consider, we will see qualitatively similar behavior below.
In general this tendency for the small-scale power to be increased over the
theoretical prediction for plain shear needs to be considered before
attributing ``excess small-scale power'' to additional physical effects
e.g.~intrinsic galaxy alignments.  A signature of the effect might be the
difference in the sensitivity of the various statistics that we compute here.

\begin{figure}
\begin{center}
\resizebox{5.0in}{!}{\includegraphics{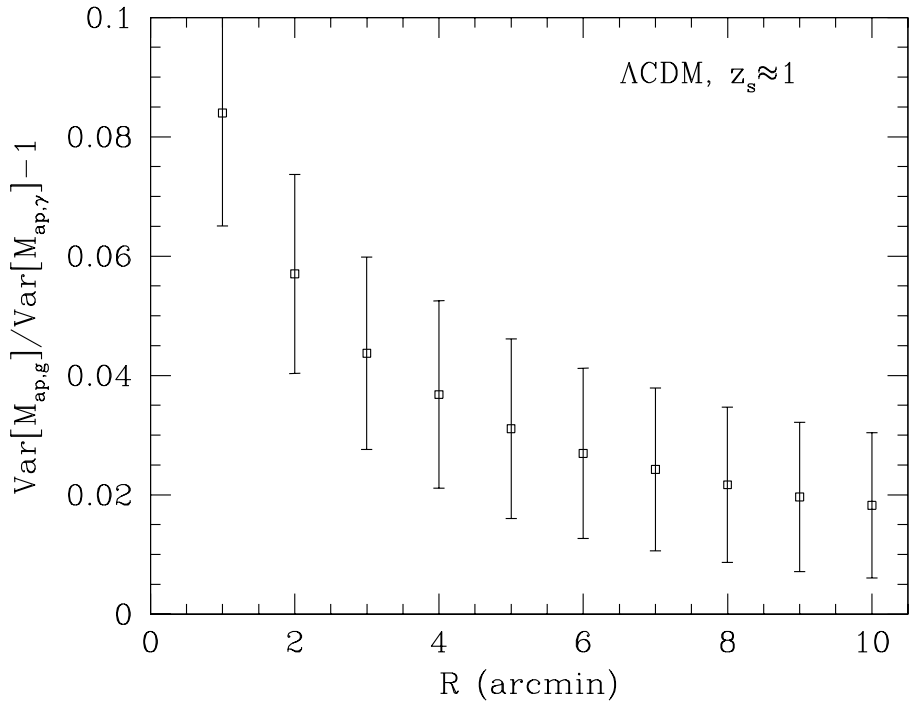}}
\end{center}
\caption{The relative difference of the $M_{\rm ap}$ variance computed for
plain shear and reduced shear.  The points show the mean difference, averaged
over 112 maps, each $3^\circ\times 3^\circ$, while the error bars indicate
the standard deviation of the ratio.}
\label{fig:mapvar}
\end{figure}

Another popular measure of the power is the aperture mass \citep{Map}.
This is a scalar quantity which can be derived from an integral over the shear
\begin{equation}
  M_{\rm ap}(\theta_0;R) \equiv
  \int d^2\theta\ Q(|\vec{\theta}|;R)\gamma_T(\vec{\theta}+\vec{\theta}_0)
\end{equation}
where $\gamma_T$ is the tangential shear as measured from $\theta_0$ and
$Q$ is a kernel.
We have chosen the $\ell=1$ form for definiteness, so
\begin{equation}
  Q(r=x/R;R) = \frac{6}{\pi R^2}\ r^2(1-r^2)\ {\rm for}\ r\le 1
\end{equation}
and $Q$ vanishes for $r>1$.
The results for the aperture mass variance are shown in
Figure \ref{fig:mapvar}.  Note that the correction is quite large on
small scales, but drops rapidly on scales above a few arcminutes.

\begin{figure}
\begin{center}
\resizebox{5.0in}{!}{\includegraphics{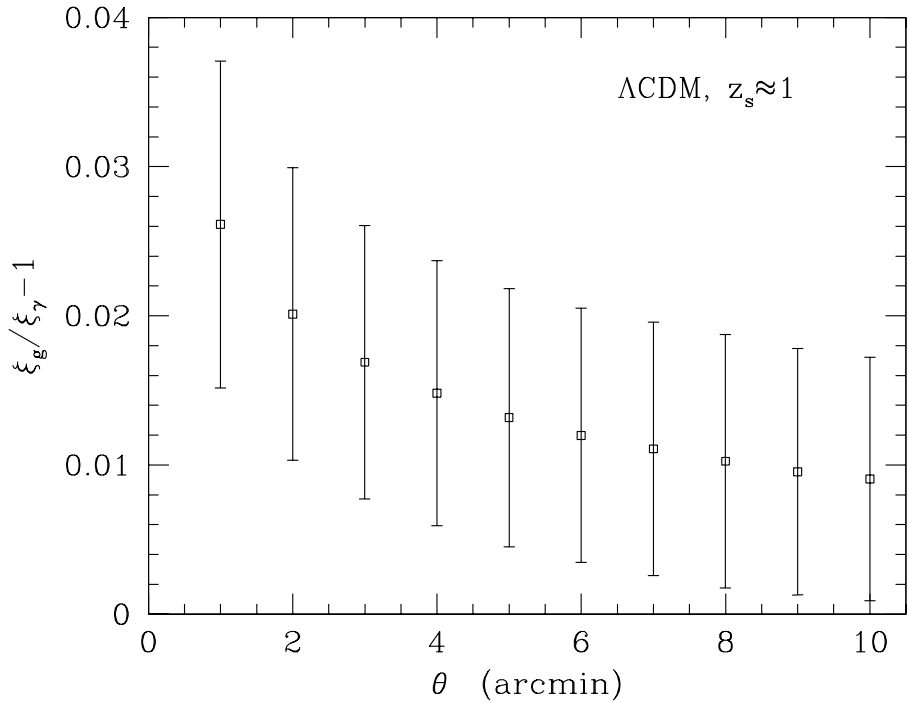}}
\end{center}
\caption{The relative difference of $\xi(r)$ computed for plain shear and
reduced shear.  The points show the mean difference, averaged over 112 maps,
each $3^\circ\times 3^\circ$, while the error bars indicate the standard
deviation of the difference.  The effect on
$\langle\gamma_\times\gamma_\times\rangle$ is smaller and is omitted.}
\label{fig:xip}
\end{figure}

Finally we show results for the 2-point correlation function of the shear
in Figure \ref{fig:xip}.  This is the best behaved 2-point statistic that
we consider, since it explicitly eliminates the contribution from small
angular scales\footnote{For this reason the 2-point correlation function is
less sensitive to small-scale observational systematics than other
quantities such as Var$[M_{\rm ap}]$.}.
We define $\xi=\langle\gamma_+\gamma_+\rangle$ where $\gamma_+$ is (minus)
the component $\gamma_1$ of the shear in the rotated frame whose
$\hat{x}$-axis is the separation vector between the two points being
correlated.  The component at $45^\circ$ is called $\gamma_\times$.
We find that the correction to the correlation function is quite small on
the scales shown here.  The correction to the other correlation function,
$\langle\gamma_\times\gamma_\times\rangle$, is even smaller than for $\xi$.

\begin{figure}
\begin{center}
\resizebox{5.0in}{!}{\includegraphics{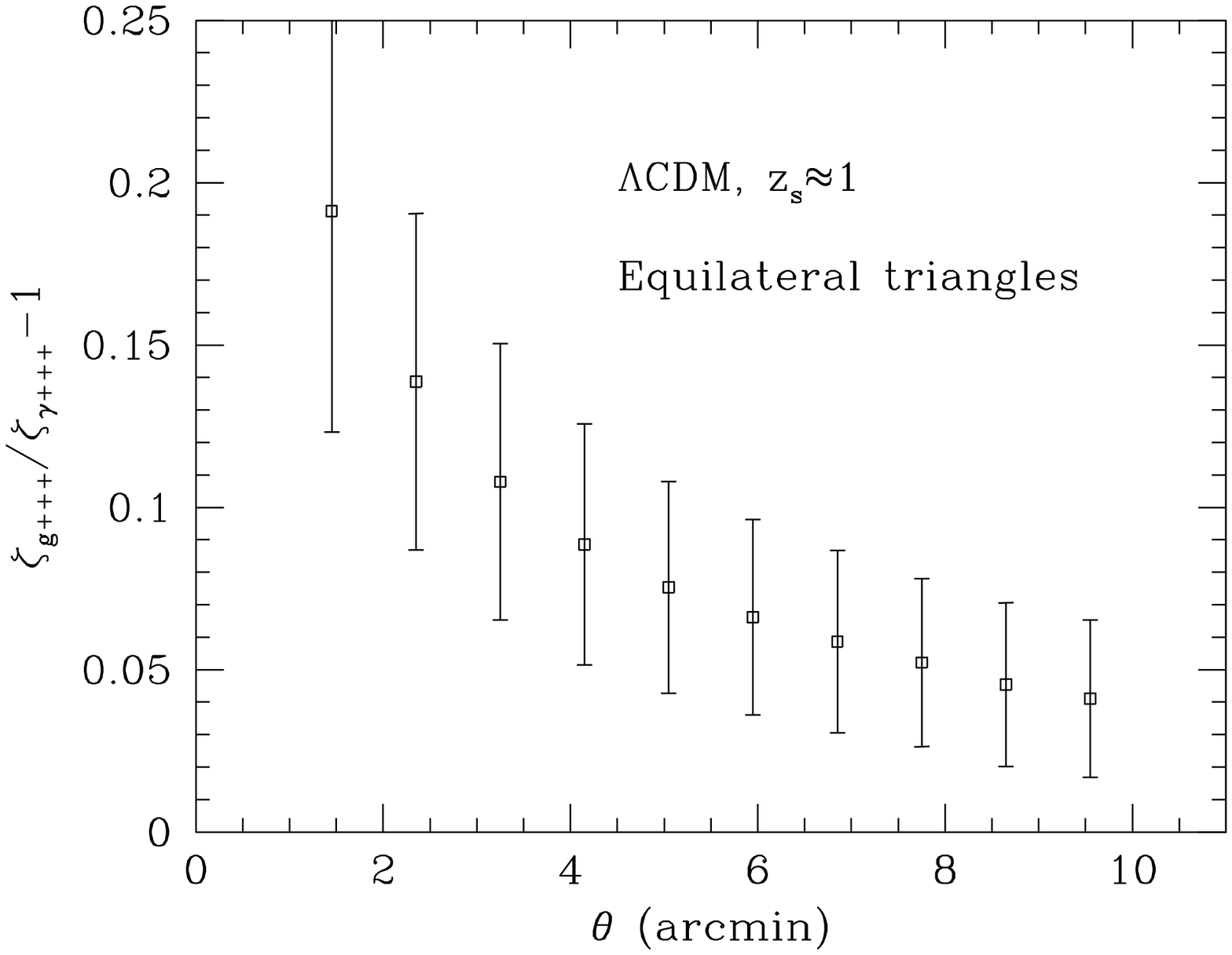}}
\end{center}
\caption{The relative difference of the 3-point correlation function,
$\zeta_{+++}$, for equilateral triangle configurations computed for plain
shear and reduced shear.  The points show the mean difference, averaged
over 32 maps, each $3^\circ\times 3^\circ$, while the error bars indicate
the standard deviation.}
\label{fig:equ}
\end{figure}

It is not too surprising that $\xi$ receives smaller corrections than
Var$[M_{\rm ap}]$ when we recall that $M_{\rm ap}$ probes smaller scales
than $R$ by a factor of roughly $3$ \citep{Map}.
A smaller piece of the difference is that $M_{\rm ap}$ can be expressed
as an integral over $\xi$ extending all the way to zero lag\footnote{The
variance of any convolution of $\kappa$ must have weight at zero lag.},
making $M_{\rm ap}$ slightly more sensitive to the difference between $g$
and $\gamma$ on small scales.
A similar argument holds for Var$[|g|_R]$, with the effect being more
pronounced.
A reduction in the sensitivity to small scale power and to shot-noise
could be obtained from differencing these measures between scales, with
an associated loss of power.

These are all of the 2-point functions commonly derived from shear
data -- the power spectrum has only been derived by two groups
\citep{descart,combo17} despite being almost ubiquitous in forecasts
of the potential of future lensing experiments
\citep[e.g.][]{Hui99,Hut01,BenBer,Hu01,WeiKam,MunWan,WL3,TakWhi,HutTak}.

\subsection{Three point statistics}

Since the lensing maps are non-Gaussian there is information
beyond\footnote{This information is not, however, independent.  We find a
strong correlation between the amplitude of the 2- and 3-point functions
on scales of several arcminutes in our maps.} the 2-point statistics.
We show in Figure \ref{fig:equ} the dominant 3-point function for equilateral
triangles as a function of scale.  The 3-point function is
$\langle\gamma_+\gamma_+\gamma_+\rangle$ with $\gamma_+$
defined with respect to a line joining the triangle center to each vertex
\citep{TakJai,ZalSco,SchLom}.
In the weakly non-Gaussian limit one can argue that the corrections to the
3-point function from considering reduced shear could be very large.
Counting powers of our small parameter, $\kappa$ or $\gamma$, the lowest
order contribution vanishes in the Gaussian limit.  Thus the first
non-vanishing term is of order $\kappa^4$ \citep{Map,DodZha}.
The first correction from the reduced shear also comes in at order
$\kappa^4$.  However we see that the difference is not significantly larger
than in the case of the 2-point functions on the scales shown here.

\begin{figure}
\begin{center}
\resizebox{5.0in}{!}{\includegraphics{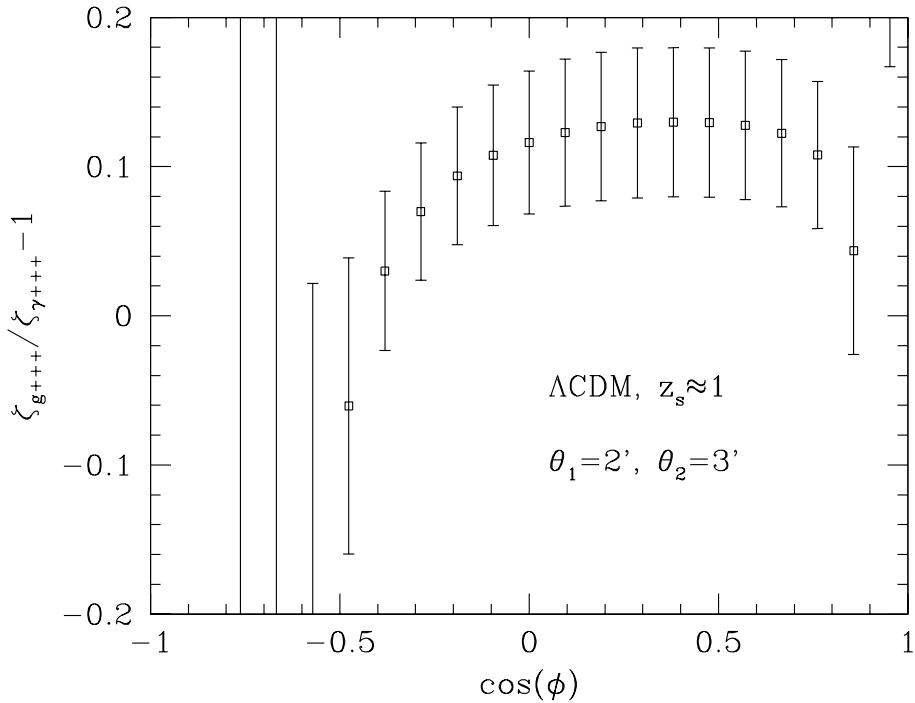}}
\end{center}
\caption{The relative difference of the 3-point correlation function for
one triangle size, $\zeta_{+++}(\theta_1=2', \theta_2=3')$, computed for
plain shear and reduced shear.  The points show the mean difference,
averaged over 32 maps, each $3^\circ\times 3^\circ$, while the error bars
indicate the standard deviation.}
\label{fig:tpt}
\end{figure}

We also show, in Figure \ref{fig:tpt}, how the configuration dependence
of the 3-point function is modified.  The 3-point function for triangles
with sides $\theta_1=2'$ and $\theta_2=3'$ is plotted as a function of
the cosine of the included angle.
We have chosen here a triangle with no special symmetries in order to
illustrate a ``typical'' case.
Note that the correction is dependent on angle, showing that the shear
and reduced shear have different shape dependence which should be
taken into account in cosmological inferences
\citep[e.g.][]{HoWhi,DolJaiTak}.

\section{Conclusions}

Deflection of light rays by gravitational potentials along the line of
sight introduces a mapping between the source and image plane.  The
Jacobian of this mapping defines the shear and convergence as a function
of position on the sky.  In the absence of size or magnification information
neither the shear nor the convergence is observable, rather the combination
$g=\gamma/(1-\kappa)$ is.  Since on small scales $\kappa$ can be
non-negligible this introduces complications in predicting the observables
of weak lensing.
We have illustrated specifically the effect of using $g$ rather than
$\gamma$ on a number of commonly used statistics of weak lensing.  We
found that the correlation function is least affected, as expected since
it specifically eliminates small-scale information.  The variance of the
shear amplitude, smoothed on a scale $R$, is the most dramatically
affected, and the amplitude of the effect declines very slowly with
increasing smoothing scale.

Existing lensing experiments derive most of their cosmological constraints
{}from angular scales of a few arcminutes to a few tens of arcminutes.
The corrections described here are below the current statistical errors,
and so should not affect existing constraints.  The next generation of
experiments may have sufficient control of systematic errors, and sufficient
statistics, that this effect will need to be included in the analysis.
Note that even for the correlation function the difference between the
commonly predicted plain shear result and the reduced shear value is
comparable to the level of systematic control that we need to achieve to
study dark energy!
If we are to reap the science rewards of future, high precision measurements
of cosmic shear we will need to be able to include this effect in our
theoretical predictions.

M.W. thanks M.~Takada for comparison of results on the shear 2-point
function and D.~Huterer, B.~Jain and P.~Zhang for helpful comments.
The simulations used here were performed on the IBM-SP at NERSC.
This research was supported by the NSF and NASA.

\end{document}